\documentclass[twocolumn,,showpacs,amssymb,amsmath,aps,pra]{revtex4-1}
\newcommand{\half}{\mbox{$\frac{1}{2}$}}
\usepackage{graphics}
\usepackage{bm}
\begin{document}
\title{Thermal entanglement in fully connected spin systems and its RPA
description}
\author{J.M.\ Matera, R.\ Rossignoli, N.\ Canosa}
\affiliation{Departamento de F\'{\i}sica-IFLP,
Universidad Nacional de La Plata, C.C.67, La Plata (1900), Argentina}
\begin{abstract}

We examine the thermal pairwise entanglement in a symmetric system of $n$ spins
fully connected through anisotropic $XYZ$-type couplings embedded in a
transverse magnetic field. We consider both the exact evaluation together with
that obtained with the static path + random phase approximation (RPA) and the
ensuing mean field + RPA. The latter is shown to provide an accurate analytic
description of both the parallel and antiparallel thermal concurrence in large
systems. We also analyze the limit temperature for pairwise entanglement, which
is shown to increase for large fields and to decrease logarithmically with
increasing $n$. Special finite size effects are as well discussed.
\end{abstract}
\maketitle

\section{Introduction}
Quantum entanglement, one of the most fundamental and intriguing features of
quantum mechanics, is well recognized as an essential resource for quantum
information processing and transmission \cite{NC.00,Be.93,BD.00}. It has
recently acquired an important role also in many-body and condensed matter
physics \cite{ON.02,V.03,T.04,AOFV.08}, where it provides a new perspective for
analyzing  quantum correlations and quantum phase transitions, as well as in
other fields like the foundations of statistical mechanics \cite{PSW.06}. The
study of entanglement in interacting spin models has in particular attracted
much interest \cite{ON.02,V.03,T.04,AOFV.08,ABV.01,GKVB.01,W.02,
VPM.04,VPA.04,DV.05,V.06,AK.05,CR.06}, since they provide a basic scalable
qubit representation suitable for implementing quantum processing tasks and are
at the same time able to capture the main features of diverse physical systems.
Some of these models can in addition be exactly solved for any size, providing
hence a suitable scenario for testing the accuracy of approximate descriptions.

An example is that of a symmetric array of $n$ fully connected spins (simplex)
with anisotropic $XYZ$ type couplings embedded in a uniform transverse magnetic
field. This is a solvable yet non-trivial model which exhibits a quantum phase
transition at $T=0$, whose Hamiltonian is formally equivalent to that of the
well-known Lipkin-Meshkov-Glick (LMG) model \cite{LMG.65}. It has attracted
renewed interest in recent years, having been used to describe diverse physical
systems such as Josephson junction arrays \cite{MSS.01} and two mode Bose
Einstein condensates \cite{CLMZ.98}. Its zero temperature entanglement
properties were analyzed in detail in refs.\ \cite{VPM.04,VPA.04,DV.05,V.06},
where it was shown in particular  that the pairwise concurrence, a measure of
the entanglement between two spins \cite{W.98}, exhibited a rich behavior when
properly scaled, with a cusplike maximum at the critical field in the
ferromagnetic case and a smooth decrease for large fields  \cite{DV.05}.

In this work we will examine the {\it thermal} pairwise entanglement in this
system, together with its description in the framework of the mean field +
random phase approximation (RPA) derived from the path integral representation
of the partition function \cite{HS.58,CMR.07}. Our aim is twofold. First, we
want to determine its thermal behavior and stability, a relevant aspect in
physical realizations. Secondly, given the complexity of determining the
entanglement properties in general interacting many-body systems at finite
temperature, we want to examine the extent to which its main features can be
captured by a general tractable method like the RPA, which takes into account
just small amplitude quantum fluctuations around the mean field. We will show
that the present RPA treatment provides, for anisotropic couplings, an accurate
analytic description of both the parallel and antiparallel thermal pairwise
entanglement in large systems, generalizing the results of \cite{CMR.07} for
the $XXZ$ case (where entanglement is just antiparallel). The limit temperature
will be shown to decrease only logarithmically with increasing $n$ at all
fields, for the standard $1/n$ scaling of coupling strengths, and to exhibit a
different field dependence in the parallel and antiparallel sectors. In
particular, it {\it increases} for large increasing fields, despite the
decrease of the $T=0$ concurrence (and at variance with the behavior in the
$XXZ$ case \cite{CMR.07}), entailing just a finite separable field window at
any temperature.

Section II describes the model and its exact partition function and
concurrence, together with their evaluation in the static path and mean field +
RPA treatments and the asymptotic expressions. Section III discusses in detail
the exact numerical and approximate results in finite systems. Finally,
conclusions are drawn in IV.

\section{Formalism}
\subsection{Exact partition function and concurrence}
We will consider $n$ qubits or spins $1/2$ coupled through an anisotropic full
range $XYZ$ Heisenberg interaction in a transverse magnetic field $b$. The
Hamiltonian reads
\begin{eqnarray}
H&=&b\sum_{i=1}^ns^z_i-\frac{1}{n}\sum_{i\neq j}^n
(v_x s_x^i s_x^{j}+v_y s_y^{i}s_y^{j}+v_z s_z^is_z^j)\nonumber\\
&=&bS_z-\frac{1}{n}\sum_{\mu=x,y,z}v_\mu (S_\mu^2-\frac{n}{4})\,,
 \label{H}\end{eqnarray}
where $s^i_\mu$ denotes the spin component at site $i$ (in units of $\hbar$)
and $S_\mu=\sum_{i=1}^n s^i_\mu$ the total spin components. The $1/n$ scaling
of coupling strengths ensures that all intensive energies remain finite for
$n\rightarrow\infty$ and finite $v_\mu$. For $T>0$ the total spin $S^2=\sum_\mu
S_\mu^2$ is no longer fixed, so that all terms in (\ref{H}) are independent.
Nonetheless,  without loss of generality we can assume $|v_y|\leq |v_x|$ and
$b\geq 0$. We will consider here the attractive case $v_x>0$ (with $|v_y|\leq
v_x)$ where the ground state will have maximum spin $S=n/2$.

Since $H$ is completely symmetric and commutes with both $S^2$ and the
$S_z$-parity  $P=\exp[i\pi (S_z+n/2)]$ (global phaseflip) the partition
function at temperature $T=\beta^{-1}$ (we set Boltzmann constant $k=1$) can be
written as
\begin{equation}
Z={\rm Tr}\exp[-\beta H]=\sum_{S=\delta_n}^{n/2}Y(S)\sum_{\nu=\pm,k}
 e^{-\beta E_{Sk^\nu}}\,,\label{Z}
\end{equation}
where $Y(S)=(^{\;\;\;\;n}_{n/2-S})-(^{\;\;\;\;\;n}_{n/2-S-1})$, with
$Y(n/2)=1$, is the multiplicity of states with total spin $S$,  such that
$\sum_{S=\delta_n}^{n/2}Y(S)(2S+1)=2^n$ [$\delta_n=0$ ($\half$) for $n$ even
(odd)]  and $E_{S k^\nu}$ are the eigenvalues of $H$ with total spin $S$ and
parity $\nu$ [$k=1,\ldots,S+\half+\nu(\half-\delta_n)$]. It should be noticed
that in the fermionic realization \cite{LMG.65}, the multiplicities $Y(S)$
would be different (the total number of states in the half-filled fermionic
system is $(^{2n}_{\;n})$ instead of $2^n$).

The pairwise entanglement at $T>0$ is determined by the reduced two-spin
density matrix $\rho_{ij}={\rm Tr}_{n-\{ij\}}\rho$ ($i\neq j$), where
$\rho=Z^{-1}\exp[-\beta H]$ is the global thermal density. $\rho_{ij}$ will be
entangled if it cannot be written as a convex combination of product densities
\cite{W.89}, i.e., if $\rho_{ij}\neq \sum_{\alpha}q_\alpha
\rho_i^\alpha\otimes\rho_j^\alpha$, with $q_\alpha>0$, being separable
otherwise. The amount of pairwise entanglement can be measured through the
entanglement of formation $E_{ij}$ \cite{Be.96},  which in the case of two
qubits can be evaluated as \cite{W.98} $E_{ij}=-\sum_{\nu=\pm} q_\nu\log_2
q_\nu$, with $q_\pm=\half(1\pm\sqrt{1-C_{ij}^2})$ and $C_{ij}$ the {\it
concurrence} \cite{W.98}, itself an entanglement measure  \cite{RC.03}. Since
$E_{ij}$ is in this case just an increasing function of $C_{ij}$, with
$C_{ij}=E_{ij}=1$ $(0)$ for a maximally entangled (separable) pair, it is
equivalent to use $C_{ij}$ as measure.

In the present system $\rho_{ij}$ will be the same for any pair and will
commute with the reduced parity $\exp[i\pi(s_z^i+s_z^j+1)]$ and total spin
$\sum_\mu(s_\mu^i+s_\mu^j)^2$, being in the standard basis of the form
\[\rho_{ij}=
\left(\begin{array}{cccc}p_+&0&0&\alpha_+\\0&p_0&\alpha_-&0\\0&\alpha_-&p_0&0\\
\alpha_+&0&0&p_-\end{array}\right),\;\;
\begin{array}{l}\alpha_{\pm}=\langle s_+^is_{\pm}^j\rangle=
\alpha_x\mp \alpha_y\\
p_{\pm}={\textstyle\frac{1}{4}}+\alpha_z\pm\langle s_z\rangle\\
 p_0={\textstyle\frac{1}{4}}-\alpha_z\end{array}\]
where $s^i_\pm=s^i_x\pm is^i_y$ and ($\mu=x,y,z$)
\begin{eqnarray}
\alpha_{\mu}&\equiv&\langle s_\mu^is_\mu^j\rangle=
\frac{T}{n-1}\frac{\partial\ln Z}{\partial v_\mu}\;\;(i\neq j)\label{amu}\,,\\
\langle s_z\rangle&\equiv&\langle s_z^i\rangle
 =-\frac{T}{n}\frac{\partial\ln Z}{\partial b}\,.
  \label{sz}\end{eqnarray}
Note that $-\frac{1}{4(n-1)}\leq\alpha_\mu\leq \frac{1}{4}$ as $\langle
S_\mu^2\rangle=\frac{n}{4}+n(n-1)\alpha_\mu$. The ensuing concurrence  $C\equiv
C_{ij}$ can be expressed as $C={\rm Max}[C_+,C_-,0]$, with
\begin{eqnarray}
C_+&=&2(|\alpha_+|-p_0)=
2(\alpha_x-\alpha_y+\alpha_z-{\textstyle\frac{1}{4}})\label{Cpp}\,,\\
C_-&=&2(|\alpha_-|-\sqrt{p_+p_-})\nonumber\\&=&2(\alpha_x+\alpha_y
-\sqrt{({\textstyle\frac{1}{4}}+\alpha_z)^2-\langle s_z\rangle^2})\,. \label{Cmm}
 \end{eqnarray}
Here $C_+$ ($C_-$) denotes a concurrence of parallel (antiparallel) type
\cite{F.06}, as in Bell states
$|\!\!\uparrow\uparrow\rangle\pm|\!\!\downarrow\downarrow\rangle$
($|\!\!\uparrow\downarrow\rangle\pm|\!\!\downarrow\uparrow\rangle$). Just one
can be positive for a given $\rho_{ij}$. In the final expressions
(\ref{Cpp})--(\ref{Cmm}) we have assumed $\alpha_{\pm}>0$, valid for the
present attractive case $v_x\geq |v_y|$. Since all pairs are equally entangled,
the maximum value that can be attained by $C$ in the present system is $2/n$
\cite{KBI.00} (reached for instance in the $W$-state
$|SM\rangle=|\frac{n}{2},\frac{n}{2}-1\rangle$) implying that only the rescaled
concurrence $c=nC$ can remain finite in the thermodynamic limit $n\rightarrow
\infty$.

 \subsection{Static path + RPA}
The auxiliary field path integral representation of the partition function
(\ref{Z}) can be written as \cite{HS.58}
\begin{eqnarray}Z&=&\int D[\bm{r}]\,
{\rm Tr}\,\left[\hat{T}\exp\{-\!\int_0^\beta
\!\!\!H[\bm{r}(\tau)]d\tau\}\right]\label{Zpi}\,,\\
H(\bm{r})&=&bS_z-\bm{r}\cdot \bm{S}+\frac{1}{4}\sum_\mu (n\frac{r_\mu^2}
 {v_\mu}+v_\mu)\,,\end{eqnarray}
where $\bm{r}=(x,y,z)$,  $\hat{T}$ denotes (imaginary) time ordering and
$H(\bm{r})$ represents a linearized Hamiltonian. Normalization $\int D[\bm{r}]
\exp[-\int_0^\beta \sum_\mu \frac{n r_\mu^2(\tau)}{4v_\mu}d\tau]=1$ is assumed.
Starting from a Fourier expansion $\bm{r}(\tau)=\sum_{k=-\infty}^\infty
\bm{r}_{k}e^{i\omega_k\tau}$, $\omega_k=2\pi k/\beta$, with $D[\bm{r}]\propto
\prod_{k}d^3\bm{r}_{k}$, the static path + random phase approximation
\cite{P.91, AA.97,RC.97} [to be denoted as correlated SPA (CSPA)] preserves the
full integral over the static components $\bm{r}\equiv \bm{r}_{0}$ but
integrates over $\bm{r}_{k}$, $k\neq 0$, in the saddle point approximation, for
each value of the running static variables. This procedure takes thus into
account large amplitude static fluctuations, relevant in critical regions,
together with small amplitude quantum fluctuations, and is feasible above a low
breakdown temperature $T^*$. The final result for the present spin $1/2$ system
can be cast as
\begin{equation}Z_{\rm CSPA}=
\sqrt{\prod_\mu\frac{n\beta}{4\pi v_\mu}}\int_{-\infty}^\infty
 \!\!\!\!\!Z(\bm{r})
\!\!\,\frac{\omega(\bm{r})\sinh[\half\beta\lambda(\bm{r})]}
{\lambda(\bm{r})\sinh[\half\beta\omega(\bm{r})]}d^3\bm{r}\,,
\label{Zcspa}\end{equation}
where, defining $\bm{\lambda}=\bm{r}-\bm{b}=(x,y,z-b)$,
\begin{eqnarray}
Z(\bm{r})&=&{\rm Tr}\exp[-\beta H(\bm{r})]\nonumber\\
&=&e^{-\frac{1}{4}\beta\sum_\mu (nr_\mu^2/v_\mu+v_\mu)}
[2\cosh\half\beta\lambda(\bm{r})]^n\,,\label{Zr}\\
\lambda(\bm{r})&=&[\sum_{\mu}\lambda_\mu^2]^{1/2}\,,\label{la}\\
\omega(\bm{r})&=&[\sum_\mu\lambda_\mu^2
(1-f_{\mu'})(1-f_{\mu''})]^{1/2}\,,\label{w}\\
f_\mu&=&v_\mu\tanh[\half\beta\lambda(\bm{r})]/\lambda(\bm{r})\,,
 \label{fmu}\end{eqnarray}
with $\mu'<\mu''$, $\mu',\mu''\neq\mu$. In (\ref{Zcspa}) $Z(\bm{r})$ is a
Hartree-like partition function while the remaining factor accounts for the
small amplitude quantum corrections, with $\omega(\bm{r})$ the single
collective thermal RPA energy existing in the present system. It can be
obtained from the equation
\begin{equation}
{\rm Det}[\delta_{\mu\mu'}-
2v_\mu \sum_{\nu=\pm}s^\nu_\mu s_{\mu'}^{-\nu}\frac{p_{-\nu}-p_{\nu}}
{\varepsilon_\nu-\varepsilon_{-\nu}-\omega}]=0\,,
 \label{rpa1}\end{equation}
with $s_\mu^\nu\equiv\langle \nu|s_\mu|\!\!-\!\!\nu\rangle$,
$p_\nu=e^{-\beta\varepsilon_\nu}/\sum_\nu e^{-\beta\varepsilon_\nu}$ and
$|\nu\rangle$, $\varepsilon_\nu$ the eigenstates and eigenvalues of
$\bm{\lambda}\cdot\bm{s}$. If $v_\mu<0$, the corresponding integral should be
done along the imaginary axes and can be evaluated in the saddle point
approximation \cite{RC.97}. The elements (\ref{amu})-(\ref{sz}) become
\[\alpha_\mu=
\frac{1}{2(n-1)}\langle\frac{n r_\mu^2}{2v_\mu^2} -\frac{1}{\beta
v_\mu}-\frac{1}{2} +(\frac{2}{\beta\omega}-\coth\half\beta\omega)
 \frac{\partial\omega}{\partial v_\mu}\rangle\]
and $\langle s_z\rangle=\half\langle z\rangle/v_z$, where $\langle
\ldots\rangle$ denotes CSPA averages.

\subsection{Mean field + RPA}
For sufficiently large $n$ and away from the critical region, we may integrate
all variables $\bm{r}_k$, including $\bm{r}_0$, in the saddle point
approximation around the minimum of the free energy potential $-T\ln
Z(\bm{r})$, determined by the self-consistent equations
\begin{equation}
r_\mu=f_\mu(r_\mu-b_\mu)\,, \;\;\mu=x,y,z \label{mef}\,.\end{equation} This
leads to the mean-field+RPA (MF+RPA). For an {\it isolated} minimum at
$\bm{r}=\bm{r}_0$, we obtain
\begin{equation}
Z_{\rm MF+RPA}=\frac{Z(\bm{r_0})}{\sqrt{1-\zeta}}
\frac{\sinh\half\beta\lambda}{\sinh\half\beta\omega}\,,
 \label{Zcmfa}\end{equation}
where $\zeta=1-\frac{\lambda^2}{\omega^2}{\rm Det}
[-\frac{2v_\mu}{n\beta}\frac{\partial^2\ln Z(\bm{r})}{\partial r_\mu
\partial r_{\mu'}}]_{\bm{r_0}}$ accounts for the gaussian static fluctuations
and $\lambda\equiv\lambda(\bm{r_0})$,  $\omega\equiv \omega(\bm{r_0})$. In
(\ref{Zcmfa}),  $Z(\bm{r_0})$ is the MF partition function while the last
factor is the proper RPA correction, which represents the ratio of two
independent boson partition functions: that of bosons of energy $\omega$ to
that of bosons of energy $\lambda$.

For the present Hamiltonian  Eqs.\ (\ref{mef}) imply either $r_\mu=0$ or
$f_\mu=1$ for $\mu=x,y$. For $|v_y|<v_x$ and $v_z<v_x$, we then obtain
the following minima:  \\
a) If $|b|<b_c$  and $T<T_c(b)$, where
\begin{equation}b_c=v_x-v_z,\;\;\;
T_c(b)=\frac{v_xb/b_c}{\ln\frac{1+b/b_c}{1-b/b_c}}\,, \label{Tc}\end{equation}
the minimum corresponds to the degenerate {\it parity-breaking} solution
$\bm{r}=(\pm x,0,z)$, with $x\neq 0$. In this case $\lambda$ is determined by
the equation $f_x=1$, i.e.,
\begin{equation}\lambda=v_x\tanh\half\beta\lambda\,,\label{la1}\end{equation}
which depends just on $v_x$ and $T$ ($\lambda=v_x$ at $T=0$), while $z=-v_z
b/b_c$ (independent of $T$) and $x=\sqrt{\lambda^2-v_x^2b^2/b_c^2}$, the
constraint $\lambda>v_xb/b_c$ leading to Eq.\ (\ref{Tc}). At this solution, the
RPA energy (\ref{w}) becomes
\begin{equation}
\omega=x\sqrt{(1-f_y)(1-f_z)},\label{w1}
\end{equation}
with $f_\mu=v_\mu/v_x$, while $\zeta=\half\beta v_x/\cosh^2\half\beta\lambda$.
Note that $\omega\rightarrow 0$ for $T\rightarrow T_c(b)$ (as $x\rightarrow 0$)
or $v_y\rightarrow v_x$ (as $f_y\rightarrow 1$), implying the divergence of
(\ref{Zcmfa}) in these limits (see \cite{CMR.07} for the correct MF+RPA
treatment in the continuously degenerate XXZ case).  \\
b) For $|b|>b_c$ or $T>T_c(b)$, the  minimum corresponds to the {\it normal}
solution $\bm{r}=(0,0,z)$. In this case $\lambda=b-z$ is the positive root
of the equation \begin{equation}\lambda=b+v_z\tanh[\half\beta\lambda]\,
\label{la2}\end{equation}
with $\lambda=b+v_z$ at $T=0$. The RPA energy becomes
\begin{equation}\omega=\lambda\sqrt{(1-f_x)(1-f_y)}.\label{w2}\end{equation}
with $f_\mu=(1-b/\lambda)v_\mu/v_z$, while $\zeta=\half\beta
v_z/\cosh^2\half\beta\lambda$.  Here $\omega\rightarrow 0$ for $T\rightarrow
T_c(b)$ (as $f_x\rightarrow 1$) but remains finite for $v_y\rightarrow v_x$.
This is also the only solution for $v_z>v_x$.

The ensuing expressions for the elements (\ref{amu})-(\ref{sz}) are
\begin{eqnarray}
\alpha_\mu&=&
{\frac{1}{2(n-1)}(\frac{nr_\mu^2}{2v_\mu^2}-\frac{1}{2}+\delta_{v_\mu}),\;
\langle s_z\rangle=\frac{1}{2n}(\frac{nz}{v_z}+\delta_{b})},\nonumber\\
\delta_\eta&\equiv&\frac{\partial\lambda}{\partial \eta}\coth\half\beta\lambda
-\frac{\partial\omega}{\partial\eta}\coth\half\beta\omega
+\frac{T}{1-\zeta}\frac{\partial\zeta}{\partial\eta}\nonumber\,,
 \end{eqnarray}
with $\eta=v_{\mu},b$. The first term in $\alpha_\mu$, $\langle s_z\rangle$ is
the $O(1)$ Hartree contribution, whereas $\delta_\eta$ provides the $O(1/n)$
RPA corrections, essential for describing entanglement.

\subsection{Asymptotic expressions for the concurrence}
Full expressions for the MF+RPA concurrence are rather long and are given in
the Appendix. However, up to $O(1/n)$ terms and for sufficiently low $T$, we
obtain
\begin{eqnarray}
C_{+}&\approx& \frac{1}{n-1}(1-\frac{\omega}{v_x-v_y}
\coth\half\beta\omega)-2e^{-\beta v_x}\label{cp1},\\
C_{-}&\approx& \frac{1}{n-1}(1-\frac{v_x-v_y}
{\omega}\coth\half\beta\omega)-2e^{-\beta v_x}
\label{cm1}\end{eqnarray}
 in the symmetry-breaking phase ($|b|<b_c$), where
\begin{equation}\frac{\omega}{v_x-v_y}=\sqrt{\frac{1-(b/b_c)^2}{1-\chi}},
\;\;\;\;\;\chi=\frac{v_y-v_z}{v_x-v_z}\,,\label{cp1a}
\end{equation}
whereas in the normal phase ($b>b_c$), $C_-\leq 0$ while
\begin{eqnarray}
C_{+}&=&\frac{1}{n-1}(1-\frac{\omega}{b+v_z-v_y}
\coth\half\beta\omega)-2e^{-\beta (b+v_z)}\label{cp2},
\end{eqnarray}
with
 \begin{equation}\frac{\omega}{b+v_z-v_y}=\sqrt{\frac{b/b_c-1}{b/b_c-\chi}}
 \label{cp2a}\,.\end{equation}
We have used in (\ref{cp1})--(\ref{cp2a}) the $T=0$ values for $\lambda$ and
$\omega$, as the ensuing thermal corrections will be of order
$e^{-\beta\lambda}=O(1/n)$ for temperatures where (\ref{cp1}), (\ref{cm1}) and
(\ref{cp2}) are positive, leading then to $O(1/n^2)$ terms in $C$. Eqs.\
(\ref{cp1})--(\ref{cp2}) become increasingly accurate as $n$ increases
(coinciding for $T\rightarrow 0$ and $v_z=0$ with the expressions of ref.\
\cite{DV.05}) and can be summarized as
\begin{equation}
C_{\pm}\approx\frac{1-(\frac{\omega}{\lambda-v_y})^{\pm 1}
\coth\half\beta\omega}{n-1}-2e^{-\beta\lambda}\,,
\label{ct}
\end{equation}
with $\lambda=v_x$, $\omega=\sqrt{(1-b^2/b_c^2)(\lambda-v_y)(\lambda-v_z)}$ for
$|b|<b_c$ and $\lambda=b+v_z$, $\omega=\sqrt{(\lambda-v_x)(\lambda-v_y)}$ for
$b>b_c$,   the result for $b>b_c$  applying just for $C_+$.

For $T\rightarrow 0$ ($e^{-\beta \lambda}\rightarrow 0$,
$\coth\half\beta\omega\rightarrow 1$) $C_{\pm}$ is then fully determined for
large $n$ by the scaled field $b/b_c$ and the anisotropy $\chi$. For $0<\chi<1$
(i.e., $v_z<v_y<v_x$), $C_-$ ($C_+$) will be positive for $|b|<b_s$ ($>b_s$),
where
\begin{equation}b_s=b_c\sqrt{\chi}\label{bs}\,,\end{equation}
is the {\it factorizing field} \cite{K.87,DV.05}, where the system possesses a
{\it separable} ground state. Accordingly, at $T=0$ both $C_{\pm}$ {\it vanish}
at $b=b_s$, $C$ being antiparallel for $|b|<b_s$ and parallel for $|b|>b_s$. On
the other hand, if $\chi\leq 0$ ($v_y\leq v_z<v_x$) or $\chi>1$ ($v_y<v_x<v_z$,
in which case there is no symmetry-breaking phase) $C$ is always parallel at
$T=0$. It is also seen from (\ref{cp2})--(\ref{cp2a}) that at $T=0$, $C_+$
remains positive for arbitrarily large fields, with $(n-1)C_+\approx
\half(1-\chi)b_c/b$ for $b\gg b_c$.

{\it Thermal effects}: Away from $b_c$ and the $XXZ$ limit, the main thermal
effect in Eqs.\ (\ref{cp1})--(\ref{cp2}) will arise from the  exponential term
$-2e^{-\beta\lambda}$, which stems in MF+RPA from the Hartree contribution to
$\alpha_\mu$ and $\langle s_z\rangle$ ($\approx \frac{1}{4}r_\mu^2/v_\mu^2$ and
$\half z/v_z$; if just this contribution is kept, Eqs.\
(\ref{Cpp})--(\ref{Cmm}) lead to $C_{\pm}=\half(\tanh^2\half\beta\lambda-1)
\approx -2e^{-\beta\lambda}$ for $\beta\lambda\gg 1$). It represents the effect
of the temperature induced decrease of the total spin average $\langle
S^2\rangle\approx \frac{1}{4}n^2\tanh^2\beta\lambda/2$. In the exact result it
arises from the lowest state of the $S=n/2-1$ multiplet, which has excitation
energy $\approx \lambda$ (see Fig.\ \ref{f4} in next section) and multiplicity
$n-1$ in Eq.\ (\ref{Z}).

The RPA thermal factor $\coth\half\beta\omega$ cannot, however, be neglected
(i.e., replaced by 1) in (\ref{cp1})--(\ref{cp2}), particularly for $b$ close
to $b_c$ or $\chi$ close to $1$, as $\omega$ is lower than $\lambda$ (for
$v_{\mu}> 0$) and vanishes for $b\rightarrow b_c$ or $v_y\rightarrow v_x$. In
the exact result it represents essentially the effect of the excited states
{\it within} the $S=n/2$ multiplet, whose excitation energies have an
approximate harmonic behavior (i.e., $\Delta E\approx k\omega$, $k=0,1,\ldots$;
see Fig.\ \ref{f4}). With this factor, Eq.\ (\ref{cm1}) correctly reduces for
$v_y\rightarrow v_x$ and up to $O(1/n)$  terms, to the asymptotic result for
the XXZ case \cite{CMR.07},
\begin{equation}
C_-\approx \frac{1}{n-1}(1-\frac{2T/b_c}{1-(b/b_c)^2})-2e^{-\beta v_x}\,,
\;\;\;\;(v_x=v_y)
\end{equation}
while for $b\rightarrow b_c$, Eqs.\ (\ref{cp1}) and (\ref{cp2}) converge  to
\begin{equation}
C_{+}\approx \frac{1}{n-1}(1-\frac{2T}{v_x-v_y})
 -2e^{-\beta v_x}\,,\;\;\;\;(b=b_c)\,.\label{Cbc}\end{equation}
Hence, in these regions the concurrence will initially exhibit an almost {\it
linear} decrease with increasing $T$ before the exponential term becomes
appreciable, as a consequence of the low excitation energy of the $S=n/2$
states.

In any case, for sufficiently large $n$, the concurrence will decrease
monotonously with increasing $T$, with $C_{\pm}$ vanishing  at a limit
temperature $T_L^{\pm}$ that will decrease {\it logarithmically} with
increasing $n$, as implied by Eq.\ (\ref{ct}):
\begin{equation}
T_L^{\pm}\approx\frac{\lambda}{\ln\frac{2(n-1)}{1-(\frac{\omega}
{\lambda-v_y})^{\pm 1}\coth\half\beta\omega}}  ,\label{TL1}
 \end{equation}
which is actually a transcendental equation for $T_L^{\pm}$. Both $T_{L}^{\pm}$
vanish (logarithmically) for $b\rightarrow b_s^{\pm}$, with $T_L^-$ decreasing
and $T_L^+$ {\it increasing} with increasing field (and $T_L^+(b)$ developing a
slope discontinuity at $b_c$). The increase of $T_L^+$ with increasing $b$ {\it
persists for $b\gg b_c$}, where
\begin{equation}
T_L^+\approx \frac{b+v_z}{\ln\frac{4(n-1)b/b_c}{1-\chi}} ,
\label{TLa}
\end{equation}
implying that at any fixed $T$ parallel entanglement can be induced by
increasing the field (for $b\gg b_c$, $\lambda$ and $\omega$ become
proportional to $b$, the system approaching then the entangled ground state as
$b$ increases). The same behavior was observed in the limit temperatures for
non-zero global negativities in small anisotropic systems \cite{CR.06}.

At fixed low $T$, the main thermal effect for $0<\chi<1$ is thus the appearance
of a {\it separable window} $b_L^-\leq |b|\leq b_L^+$ (instead of a separable
point) where $C_{\pm}=0$, with
\begin{equation}
b_L^{\pm}\approx b_c\sqrt{1-(1-\chi) [\tanh\half\beta\omega(1-2(n-1)
 e^{-\beta v_x})]^{\pm 2}}
 \label{bL}\end{equation}
(valid for $T<T_L^+(b_c)$ for $b_L^+$ and $T<T_L^-(0)$ for $b_L^-$). Its width
increases then with increasing $n$ or $T$, with $b_L^{\pm}\approx b_s[1\pm
2(\chi^{-1}-1)(n-1)e^{-\beta v_x}]$ for $ne^{-\beta v_x}\ll 1$. For
$T>T_L^-(0)$  the separable window will extend through $b=0$ ($C_{\pm}=0$ for
$|b|<b_L^+$).

\section{Comparison with exact results in finite systems}

\begin{figure}[t]

\centerline{
\hspace*{-0.35cm}
\scalebox{.75}{\includegraphics{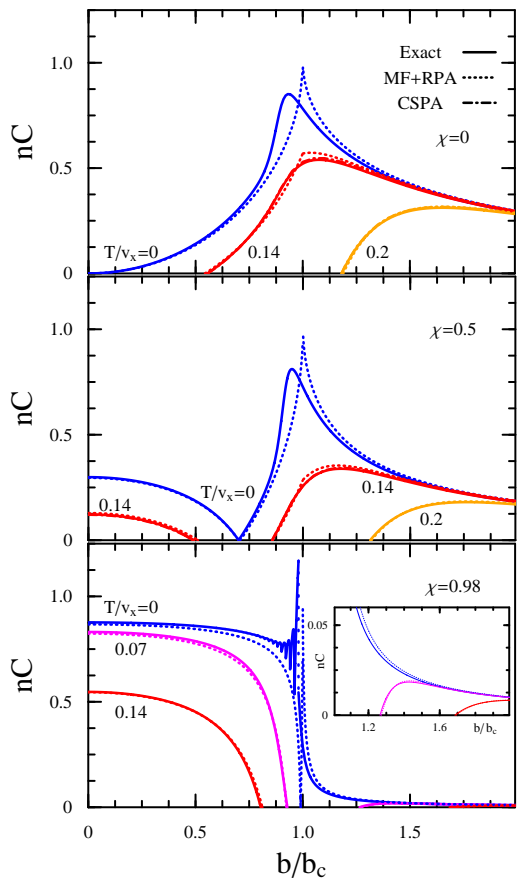}}}

\caption{(Color online) Scaled concurrence as a function of the magnetic field
$b$ for $n=100$ spins coupled through a full range $XY$ interaction with
anisotropies $\chi=v_y/v_x=0$  (top), $0.5$ (center) and $0.98$ (bottom), at
different temperatures. Exact and asymptotic mean field+RPA (Eqs.\
(\ref{cp1})--(\ref{cp2})) results are depicted, together with those of CSPA
(Eq.\ \ref{Zcspa}) for $T>0$ (almost undistinguishable from the exact ones).
$C$ is parallel (antiparallel) for $b>b_s$  ($<b_s$), with $b_s$ the
factorizing field (\ref{bs}). The inset depicts the (parallel) concurrence
reentry for $b>b_c$ at $\chi=0.98$.} \label{f1}
\end{figure}

Typical results for the magnetic behavior of the concurrence at finite
temperatures are shown in Fig.\ \ref{f1} for the $XY$ case ($v_z=0$) with
$n=100$ spins and different anisotropies. It is first seen that MF+RPA results
obtained with the asymptotic expressions (\ref{cp1})--(\ref{cp2}) are very
accurate except in the vicinity of the critical field, improving as $T$
increases. The full CSPA results further improve those of the MF+RPA in the
critical region for not too low $T$, being practically undistinguishable from
the exact ones at the finite temperatures considered.

The top panel corresponds to the Ising case $v_y=0$, where the concurrence is
always parallel. At $T=0$ it smoothly increases from $0$ as $b$ increases,
having a maximum near $b_c$, while for $T>0$ it becomes non-zero only above a
{\it threshold} field $b_L^+$ (Eq.\ (\ref{bL}) for $T<T_L^+(b_c)$). In the
central panel ($\chi=0.5$) we may appreciate the vanishing of the concurrence
at the factorizing field $b_s\approx 0.71 v_x$ at $T=0$, where it changes from
antiparallel to parallel.  This point evolves into a separable window as $T$
increases, which extends through $b=0$ for $T>T_L^-(0)\approx 0.15 v_x$.

\begin{figure}[t]
\centerline{\hspace*{-0.4cm}\scalebox{.75}{\includegraphics{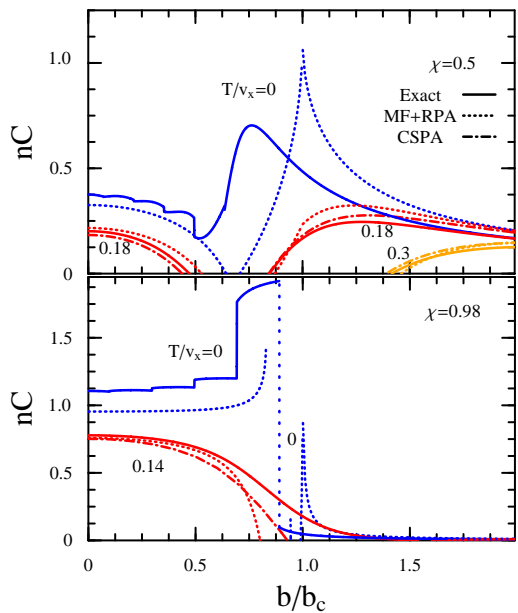}}}

\caption{(Color online) Magnetic behavior of the concurrence for a chain with
$n=10$ spins at different temperatures, for anisotropies $\chi=0.5$ (top) and
$\chi=0.98$ (bottom). In the latter the antiparallel concurrence increases with
increasing field at very low $T$,  as in the $XXZ$ case (see text).} \label{f5}
\end{figure}

The bottom panel depicts the behavior close to the $XXZ$ limit. In this case
the exact $T=0$ concurrence $C_-$ displays an oscillatory behavior as
$b\rightarrow b_s$ from below, as in the $XXZ$ chain \cite{CMR.07}, which
reflects the ground state spin parity transitions and which is not reproduced
by MF+RPA (see however discussion of Fig.\ \ref{f5}). Nonetheless, as $T$
increases the oscillations become rapidly washed out and the asymptotic MF+RPA
result becomes again accurate, correctly reproducing the exact concurrence at
$T/v_x=0.07$ and $0.14$, {\it including} the  reentry of the parallel
concurrence that takes place for high fields. The thermal RPA factor
$\coth\half\beta\omega$ is here essential for the accuracy as $\beta w$ is
small ($\omega/v_x\alt 0.14$ for $b<b_c$).

\begin{figure}[t]
\centerline{\hspace*{-0.5cm}\scalebox{.75}{\includegraphics{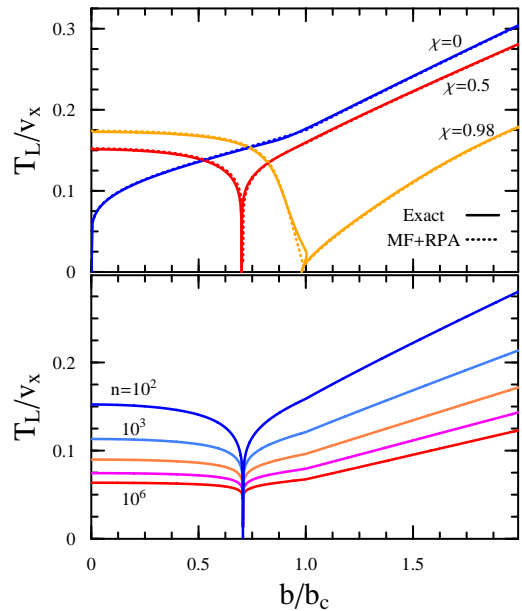}}}

\caption{(Color online) Top: Limit temperatures for pairwise entanglement $T_L$
as a function of the magnetic field $b$ for  $n=100$ spins at the same
anisotropies of Fig.\ \ref{f1}, according to exact and MF+RPA results (Eq.\
(\ref{TL1})). They vanish at the factorizing field $b_s$. Regions below the
limit temperature have finite pairwise entanglement, of antiparallel (parallel)
type if $b<b_s$ ($>b_s$). Bottom: Limit temperatures for increasing number of
spins at $\chi=0.5$ ($n=10^k$, $k=2,\ldots,6$).} \label{f2}
\end{figure}

Let us mention that for $\chi\in(0,1]$, the ground state, which has definite
spin parity $P=\pm 1$, exhibits $n/2$ transitions $\pm\rightarrow\mp$ as $b$
increases from $0$, the last one at the factorizing field $b_s$. The ground
state concurrence changes from antiparallel to parallel just at this last
transition. These transitions are, however, appreciable only for $\chi$ close
to 1 (and hence $b_s$ close to $b_c$) or for small sizes, as otherwise the
ground states of both parity sectors are practically degenerate and the
concurrence is nearly the same in both states (see Fig.\ 4) as well as in their
mixture. Another consequence of parity conservation is that the exact side
limits of $C_{\pm}$ at $b=b_s$ {\it are actually  non-zero} and different in
finite chains ($nC_{\pm}\rightarrow \delta/(e^{\delta/2}\pm 1)$, with
$\delta=n(1-\chi)$ \cite{RCM.08}), being then appreciable for small finite
$\delta$. In the bottom panel we thus obtain the side limits $nC_-\approx
1.16$, $nC_+\approx 0.54$ for the exact result at $b_s\approx 0.99 b_c$, with
$C_{\pm}$ being in fact {\it maximum} at $b=b_s$.

Fig.\ \ref{f2} depicts the corresponding limit temperatures $T_L^{\pm}$, which,
remarkably, are also accurately reproduced by the asymptotic MF+RPA result
obtained from Eq.\ (\ref{TL1}). $T_L$ vanishes at $b=b_s$ but increases
$\forall$ $b>b_s$, developing thus a separable field window between the
antiparallel and parallel concurrences. The bottom panel shows the logarithmic
decrease of $T_L$ with increasing $n$ in all regions. Let us also remark that
the behavior of $T_L$ bears no relation with that of the mean field critical
temperature, Eq.\  (\ref{Tc}), which does not depend on the anisotropy $\chi$
and vanishes for $b>b_c$. For $|b|<b_c$ it is  essentially higher than $T_L$
(except for very low $n$ \cite{CR.06}), decreasing monotonously from $\half
v_x$ at $b=0$ to $0$ at $b=b_c$.

\begin{figure}[t]

\centerline{\hspace*{-0.4cm}\scalebox{.75}{\includegraphics{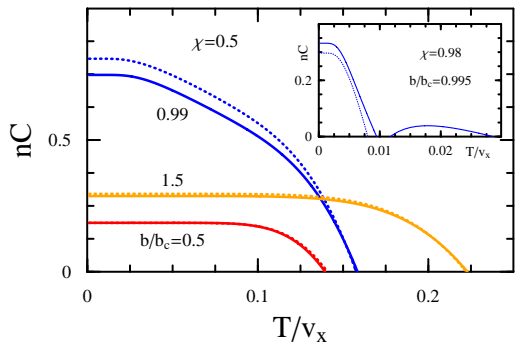}}}

\caption{(Color online) Thermal behavior of the concurrence for $\chi=0.5$ and
$n=100$ at the indicated fields. The inset depicts the non-nonotonous behavior
just above $b_s$ for $\chi=0.98$ and $n=100$. Solid lines depict exact results,
dotted lines those from MF+RPA, Eqs.\ (\ref{cp1})--(\ref{cp2}).}
 \label{f3}
\end{figure}

For $\chi=0.98$, the exact limit temperature $T_L^-$ actually exhibits a small
{\it positive slope} close to $b_s$, as seen in the top panel (not reproduced
by MF+RPA). This entails that at low finite $T$ the antiparallel concurrence
will persist in a narrow region {\it above} $b_s$, while at fixed $b$ within
this region, the thermal behavior of the concurrence will be {\it
non-monotonous}, being first parallel, vanishing and becoming then antiparallel
before extinguishing at the final $T_L^-$, as depicted in the inset of Fig.\
\ref{f3}. Roughly, for small $b-b_s>0$, it is possible to show that
$T_L^{\pm}\approx \alpha^{-1}(b-b_s)\delta e^{-\delta/2}/(1-e^{-\delta})$, with
$\alpha=\ln\coth\delta/4$, this effect being then noticeable for finite
$\delta=n(1-\chi)$.

The different thermal response of $C$ for fields below, around and above the
critical field $b_c$ can be seen in the main panel of Fig.\ \ref{f3} for
$\chi=0.5$. The more rapid decrease with increasing $T$ for  $b\approx b_c$ is
in agreement with Eq.\ (\ref{Cbc}), while the results at $b/b_c=0.5$ and $1.5$
reflect the different decrease rate (Eqs.\ (\ref{cm1})--(\ref{cp2})).

\begin{figure}[t]
\centerline{\hspace*{-0.4cm}\scalebox{.75}{\includegraphics{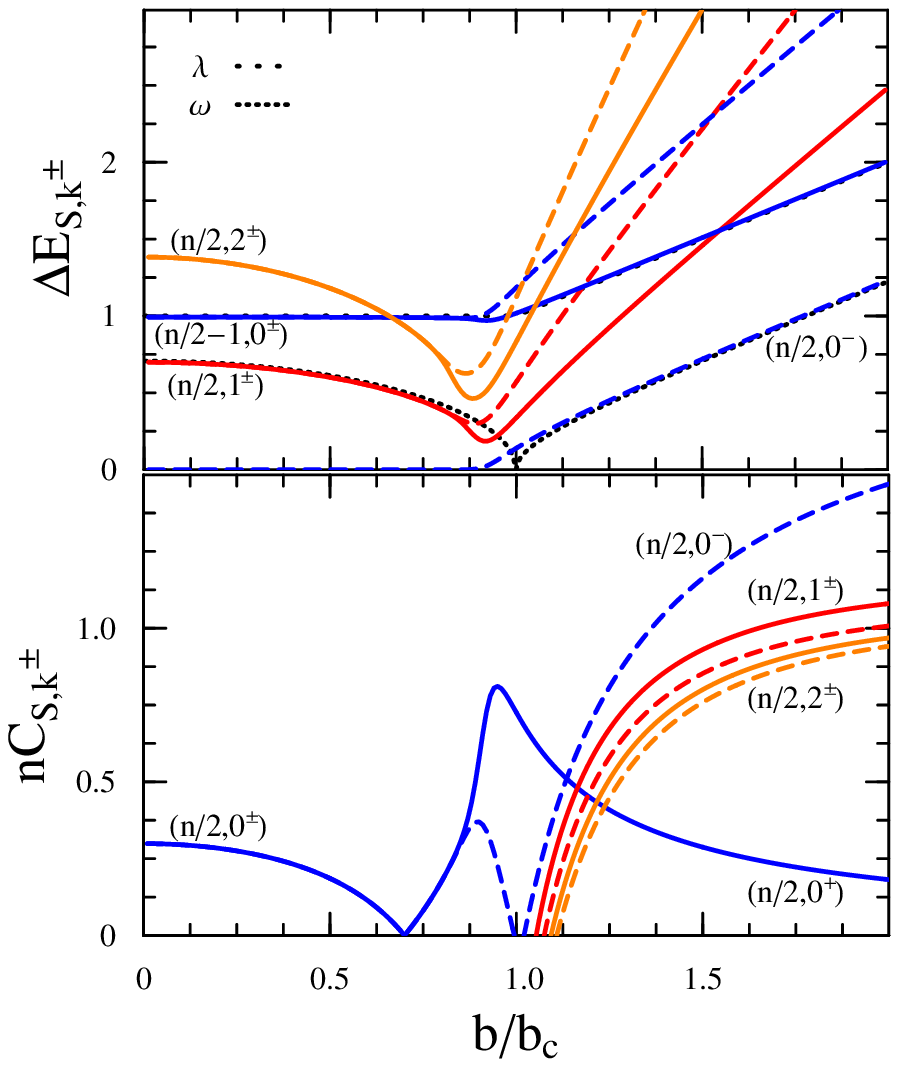}}}

\caption{(Color online) Top: Lowest excitation energies $\Delta E_{Sk\pm}\equiv
(E_{Sk\pm}-E_{0})/v_x$ for $\chi=0.5$ and $n=100$, vs.\  magnetic field $b$.
$E_{Sk^\nu}$ denotes the energy of level $k$ with total spin $S$ and parity
$\nu$, while $E_0=E_{n/2,0^+}$. Solid (dashed) lines depict levels of positive
(negative) parity, practically degenerate for $b<b_c$. The dotted lines depict
the mean field and RPA energies $\lambda$ and $\omega$ respectively. Bottom:
The corresponding concurrences, including that in the ground state. For $b<b_c$
$C$ is antiparallel (parallel) for $b<b_s$ $(>b_s)$ in both states
$(n/2,0^{\pm})$, whereas for $b>b_c$, it is parallel in the ground state
($n/2,0^+$) but  antiparallel in the other levels depicted.}
 \label{f4}\end{figure}

The origin of the distinct thermal factors in Eqs.\ (\ref{cp1})--(\ref{cp2})
can be seen in Fig.\ \ref{f4}, which depicts the excitation energies of the
lowest levels $(S,k^{\pm})$ for $\chi=0.5$. For $|b|<b_c$, corresponding levels
of opposite parity are practically degenerate. The non-vanishing excitation
energies within the maximum spin multiplet are nearly harmonic, the lowest one
practically coinciding with the RPA energy (\ref{w1}), whereas the excitation
energy of the lowest state with $S=n/2-1$ is almost $b$ independent and
coincident with $\lambda=v_x$. The parity degeneracy becomes broken in all
levels as $b$ approaches $b_c$, where the maximum spin excitations become low
and give rise to the increased thermal sensitivity (Eq.\ (\ref{Cbc})). For
$|b|>b_c$ the RPA energy represents again the lowest excitation energy in the
maximum spin multiplet, which has now negative parity, whereas $\lambda=b+v_z$
is again the excitation of the lowest $S=n/2-1$ state, which has now positive
parity.

The corresponding concurrences are shown in the lower panel. For $|b|<b_c$ and
$\chi=0.5$, $C$ is non-zero just in the degenerate ground states, being almost
coincident except for $b$ close to $b_c$  and changing both from antiparallel
to parallel at $b_s$. However, for $b>b_c$ $C$ is non-zero in all maximum spin
states, being parallel in the ground and highest states but antiparallel in the
rest. They are essentially the basic states $|S=n/2,S_z=M\rangle$ plus
perturbative corrections. For $|M|<n/2$ they are already entangled and exhibit
hence antiparallel concurrence \cite{CMR.07}, while for $|M|=n/2$ the
concurrence arises just from the corrections and is hence parallel. We also
note that the mixture of the $n-1$ states with lower spin $S=n/2-1$ has zero
concurrence at all fields (the same occurs with lower spin mixtures) so that it
can only decrease the thermal concurrence, which  arises then essentially from
the ground state, except in anomalous regions (the antiparallel reentry in the
inset of Fig.\ \ref{f3} arises from the first excited state).

Finally, Fig.\ \ref{f5} depicts results for a small chain ($n=10$), where
finite size effects become exceedingly important. The stepwise behavior of the
exact concurrence at $T=0$ is now visible already for $\chi=0.5$ (top panel),
the side limits at the exact factorizing field $b_s^{ex}=(1-n^{-1})b_s$ being
non-zero. The MF+RPA result is now less accurate at $T=0$, and leads to the
vanishing of $C_-$ at a field lower than $b_s$ (and closer to $b_s^{ex}$) if
the full expression (\ref{A2}) is used for $C_-$. Nonetheless,  MF+RPA results
rapidly improve as $T$ increases, while the CSPA, although no longer exact,
improves again results in the vicinity of $b_c$.

When the full square root in the evaluation of $C_-$ is kept (Eq.\ (\ref{A2})),
the MF+RPA is actually able to qualitatively account for a finite jump in $C$
for fields close to $b_s$ but only if $\chi$ is very  close to $1$, as seen in
the bottom panel. In this case, the MF+RPA result for $C_-$ at $T=0$ does not
decrease as $b$ increases but rather {\it increases}, in agreement with the
exact result, terminating at a final field $b_f<b_s^{ex}$ where it starts to be
complex (and is maximum). For $\chi=1-\delta/n$ and fields just below $b_c$,
i.e., $(b/b_c)^2=1-\varepsilon/n$, we actually obtain, instead of Eq.\
(\ref{cm1}), the asymptotic MF+RPA expression

\begin{eqnarray}C_-&\approx&{\textstyle \frac{1}{n}(\half\varepsilon-
\sqrt{\frac{\delta}{\varepsilon}}\coth\half\beta\omega)-2e^{-\beta v_x}}
\nonumber\\
&&{\textstyle-\frac{1}{n}[2\sqrt{\frac{\delta}{\varepsilon}}
\coth\half\beta\omega
+\frac{1}{4}\varepsilon(\varepsilon-4)]^{1/2}}
 \label{cmap}\end{eqnarray}
with $\omega=\sqrt{\varepsilon\delta}(v_x-v_z)/n$.  While for
$\varepsilon\propto n$ it reduces to Eq.\ (\ref{cm1}), for $\varepsilon<4$ it
becomes complex if $\delta$ is sufficiently small. At $T=0$, if
$\delta<\delta_c=\frac{12^3}{5^5}\approx 0.55$, Eq.\ (\ref{cmap}) becomes
complex for $\varepsilon<\varepsilon_f(\delta)\approx
2.4+\frac{5}{3}\sqrt{\delta_c-\delta}$,  with $C_-(b_f)\approx
\frac{1}{8}\varepsilon_f^2$. Note that $C_-(b_f)>1$ for $\delta\alt 0.48$, with
$C_-(b_f)\rightarrow 2/n$ for $\delta\rightarrow 0$, which is the correct
result for the $XXZ$ limit \cite{CMR.07}. In the case depicted, $\delta=0.2$.

Eq.\ (\ref{cmap}) implies as well that these effects, in particular the
increase of $C_-$ as $b\rightarrow b_s$ for $\chi$ close to $1$,   will
disappear for very low $T\propto \omega\propto b_c/n$, which is confirmed in
the exact results. One may also appreciate in the bottom panel of Fig.\
\ref{f5} the significant persistence of the antiparallel concurrence up to
$b\approx 1.5 b_c$ at $T/v_x=0.14$ (not reproduced by MF+RPA or CSPA), which is
just the same anomalous behavior discussed in Figs.\ \ref{f2}--\ref{f3},
enhanced by the smaller value of $\delta$. Nonetheless, even in this extreme
case there is a weak but non-zero revival of the parallel concurrence for high
fields $\forall$ $T$ (appreciable in the figure just for $T=0$) which is
correctly reproduced by MF+RPA.

\section{Conclusions}
We have analyzed the thermal behavior of the pairwise concurrence in a fully
connected spin system with anisotropic couplings placed in a transverse field.
For the usual $1/n$ scaling of coupling strengths, the limit temperature
decreases only logarithmically as the size $n$ increases, and decreases
(increases) for increasing field in the antiparallel (parallel) sectors, the
latter extending for arbitrarily large fields. This behavior was previously
observed for small $n$ in the temperatures limiting  global negativities, for
which the pairwise limit temperature provides a lower bound. Anisotropic arrays
become then strictly pairwise separable just within a finite field window at
any temperature, which collapses into the factorizing field at $T=0$.
Remarkably, all previous features of the pairwise entanglement can be captured
by a simple thermal MF+RPA treatment, consistently derived from the path
integral representation of the partition function, which in the present case is
able to provide a reliable analytic description of the concurrence and limit
temperature at all fields, exact in the large $n$ limit. We have also discussed
the special finite size effects exhibited by this system for small anisotropies
or sizes, whose main aspects can also be qualitatively reproduced by MF+RPA or
the full CSPA. These results suggest the possibility of describing by simple
means at least the main features of the thermal pairwise entanglement in more
complex systems, although the actual accuracy and scope of the RPA in such
situations remains to be investigated.

The authors acknowledge support from CIC(RR) and CONICET (JMM, NC) of
Argentina.

\appendix*
\section{MF+RPA Concurrence}
We provide here the expressions for the full MF+RPA concurrence derived from
Eqs.\ (\ref{amu})--(\ref{Cpp}) and (\ref{Zcmfa}). Setting $v_x=1$ and
$\tilde{b}\equiv b/b_c$, in the symmetry breaking phase we obtain
\begin{eqnarray}
C_+&=&-\frac{1-\lambda^2}{2}+\frac{1}{n-1}\{1-\frac{\omega}{1-v_y}
\coth\half\beta\omega[1+\nonumber\\
&&\frac{\zeta}{1-\zeta}\frac{(1-v_y)\lambda^2}{\lambda^2-\tilde{b}^2}]
-(\frac{\zeta}{1-\zeta})^2[1-(3-\zeta)T]\}\,,\label{A1}\\
C_-&=&\frac{\lambda^2-\tilde{b}^2}{2}+\frac{1}{n-1}\{
\frac{1-\tilde{b}^2}{2}-\nonumber\\
&&
-\frac{1-v_y}{\omega}\coth\half\beta\omega
[\frac{\lambda^2+\tilde{b^2}}{2}+\frac{\zeta}{1-\zeta}\lambda^2(1-v_z)]
-\nonumber\\
&&
-\frac{\zeta^2}{(1-\zeta)^2}[1-(3-\zeta)T]\}-\nonumber\\
&&\{[\frac{1+\tilde{b}^2}{2}+\frac{1}{n-1}(\frac{1-v_y}{\omega}
\coth\half\beta\omega
\frac{\lambda^2+\tilde{b}^2}{2}-\frac{1-\tilde{b}^2}{2})]^2\nonumber\\
&&-\tilde{b}^2(1+\frac{1-v_y}{n\omega}\coth\half\beta\omega)^2\}^{1/2}\,,
 \label{A2}\end{eqnarray}
where  $\zeta=\half\beta(1-\lambda^2)$ and
$\lambda=\tanh\half\beta\lambda$. If $\tilde{b}$ is not close to $1$, we may
expand Eq.\ (\ref{A2}) up to $O(1/n)$ as
\begin{eqnarray}
C_-&\approx& -\frac{1-\lambda^2}{2}+\frac{1}{n-1}\{1-\frac{1-v_y}{\omega}
\coth\half\beta\omega [\frac{\lambda^2-\tilde{b}^2}{1-\tilde{b}^2}\nonumber\\
&&+\frac{\zeta}{1-\zeta}(1-v_z)\lambda^2]-\frac{\zeta^2}{(1-\zeta)^2}
[1-(3-\zeta)T]\}\,.
 \label{A3}\end{eqnarray}
For $\beta\lambda\ll 1$, $\lambda\approx 1-2e^{-\beta \lambda}$, with
$1-\lambda^2\approx 4e^{-\beta\lambda}$. Eqs.\ (\ref{cp1})--(\ref{cm1}) are
then obtained from (\ref{A1}), (\ref{A3}) neglecting $\zeta$ and setting
$\lambda=1$ in the $O(1/n)$ terms.  This is correct up to  $O(1/n)$ terms for
temperatures where $C_{\pm}$ are positive, as in such a case
$e^{-\beta\lambda}$ must be $O(1/n)$.

Similarly, in the normal phase we obtain
\begin{eqnarray}
C_+&=&-\frac{1-\tanh^2\half\beta\lambda}{2}+
\frac{1}{n-1}\{1-\frac{\omega f_x}{1-f_y}
\coth\half\beta\omega+\nonumber\\
&&\half\zeta\coth\half\beta\omega\!\!\sum_{\mu=x,y} \frac{\omega f_x}{1-f_\mu}
(\frac{v_\mu}{v_z}-\frac{1}{1-\zeta})-  \nonumber\\
&&-\frac{\zeta}{(1-\zeta)}\frac{3T}{v_z}\}
 \label{A4}\end{eqnarray}
where $f_\mu=v_\mu\tanh(\half\beta\lambda)/\lambda$, $\zeta=\half\beta
v_z(1-\tanh^2\half\beta\lambda)$ and $\lambda$ determined by Eq.\ (\ref{la2}).
For $\beta\lambda\ll 1$, $1-\tanh^2\half\beta\lambda\approx
-4e^{-\beta\lambda}$. Eq.\ (\ref{A4}) leads then to Eq.\ (\ref{cp2}) up to
$O(1/n)$ for temperatures where $C_+>0$, by neglecting $\zeta$ and setting
$\lambda=b+v_z$ in the $O(1/n)$ terms.


\begin{thebibliography}{999}
\bibitem{NC.00}M.A.\ Nielsen and I. Chuang, {\it Quantum Computation and
               Quantum Information}, Cambridge Univ. Press, Cambridge, England,
                (2000).
\bibitem{Be.93} C.H.\ Bennett et al., Phys.\ Rev.\ Lett.\ {\bf 70}, 1895
    (1993).
\bibitem{BD.00}C.H.\ Bennett and D.P.\ DiVincenzo, Nature {\bf 404}, 247
 (2000).
\bibitem{ON.02} T.J.\ Osborne and  M.A.\ Nielsen, Phys.\  Rev.\ A {\bf  66},
               032110 (2002).
\bibitem{V.03} G.\ Vidal, J.I.\ Latorre, E.\ Rico, and  A.\ Kitaev,
              Phys.\ Rev.\ Lett.\ {\bf 90}, 227902 (2003).
\bibitem{T.04} T.\ Roscilde, P.\ Verrucchi, A.\ Fubini, S.\ Haas, and
V.\ Tognetti, Phys.\ Rev.\ Lett.\ {\bf 93}, 167203 (2004).
\bibitem{AOFV.08} L.\ Amico, R.\ Fazio, A.\ Osterloh and
           V.\ Vedral, Rev.\ Mod.\ Phys.\ (2008) (in press).
\bibitem{PSW.06} S. Popescu, A. Short, and A. Winter, Nature Physics {\bf 2},
  754 (2006).
\bibitem{ABV.01} M.C.\ Arnesen, S.\ Bose, and V.\ Vedral, Phys.\ Rev.\ Lett.\
               {\bf 87}, 017901 (2001).
\bibitem{GKVB.01} D.\ Gunlycke, V.M.\ Kendon, V.\ Vedral, and S.\ Bose,
              Phys.\ Rev.\ {\bf A 64}, 042302 (2001).
\bibitem{W.02} X.\ Wang, Phys.\ Rev.\ A {\bf 64}, 012313 (2001); X. Wang and P.
               Zanardi, Phys.\ Lett.\ A {\bf 301}, 301 (2002); X.\ Wang
               and Z.D.\ Wang, Phys.\ Rev.\ A {\bf 73}, 064302 (2006).
\bibitem{VPM.04} J.\ Vidal, G.\ Palacios, and R.\ Mosseri, Phys.\ Rev.\ A
 {\bf  69} 022107 (2004);
S.\ Dusuel and J.\ Vidal, Phys.\ Rev.\ Lett.\ {\bf 93}, 237204 (2004).
\bibitem{VPA.04}
J.\ Vidal, G.\ Palacios, and G. Aslangul, Phys.\ Rev.\  A {\bf 70} 062304
(2004).
\bibitem{DV.05}S. Dusuel and J.\ Vidal, Phys.\ Rev.\ B {\bf  71} 224420 (2005).
\bibitem{V.06} J.\ Vidal, Phys.\ Rev.\ A {\bf  73} 062318 (2006).
\bibitem{AK.05} M.\ Asoudeh and V.\ Karimipour, Phys.\  Rev.\ A {\bf 71},
               022308 (2005).
\bibitem{CR.06}R.\ Rossignoli and N.\ Canosa, Phys.\ Rev.\ {\bf A} 72, 012335
(2005). N.\ Canosa and  R. Rossignoli, Phys.\ Rev.\ {\bf A} 73, 022347 (2006).
\bibitem{LMG.65} H.\ J.\ Lipkin, N.\ Meshkov, and A.\ J.\ Glick, Nucl.\ Phys.\
{\bf 62}, 188 (1965).
\bibitem{MSS.01} Y.\ Makhlin, G.\ Sch\"on, and
A.\ Shnirmann, Rev.\ Mod.\ Phys.\ {\bf 73}, 357 (2001).
\bibitem{CLMZ.98}J.\ I.\ Cirac, M.\ Lewenstein,
K.\ M\o lmer, and P.\ Zoller,  Phys.\ Rev.\ A {\bf 57}, 1208 (1998).
\bibitem{W.98}S.\ Hill and  W.K.\ Wootters, Phys.\ Rev.\ Lett.\ {\bf 78}, 5022
             (1997); W.K.\ Wootters, Phys.\ Rev.\ Lett.\ {\bf 80}, 2245 (1998).
\bibitem{HS.58} R.\ L.\ Stratonovich, Dokl.\ Akad.\ Nauk SSSR {\bf 115},
1097 (1957) (Sov.\ Phys.\ Dokl.\ {\bf 2} 458 (1958));
 J.\ Hubbard, Phys.\ Rev.\ Lett.\ {\bf3}, 77 (1959).
\bibitem{CMR.07}N. Canosa, J.M.\ Matera, and R. Rossignoli,
Phys.\ Rev.\ A {\bf  76} 022310 (2007).
\bibitem{W.89} A.\ Werner, Phys.\ Rev.\ A {\bf 40}, 4277 (1989).
\bibitem{Be.96} C.H.\ Bennett, D.P.\ DiVincenzo, J.A.\ Smolin, and
               W.K.\ Wootters, Phys.\ Rev.\ A {\bf  54}, 3824 (1996).
\bibitem{RC.03} P.\ W.\ Rungta and C.\ M.\ Caves, Phys.\ Rev.\ A {\bf 67},
012307 (2003).
\bibitem{F.06}A.\ Fubini, T.\ Roscilde, M.\ Tusa, V.\ Tognetti and P.\
Verrucchi, Eur.\ Phys.\ J.\ D.\ {\bf 38}, 563 (2006).
\bibitem{KBI.00} M.\ Koashi, V.\ Buzek, and N.\ Imoto, Phys.\ Rev.\ A {\bf 62},
050302(R) (2000); W.\ D\@ur, {\it ibid} {\bf 63}, 020303 (2001).
\bibitem{P.91}G.\ Puddu, P.F.\ Bortignon, and R. Broglia, Ann.\ Phys.\ (N.Y.)
{\bf 206}, 409 (1991).
\bibitem{AA.97}H.\ Attias and Y. \ Alhassid,  Nucl. Phys.\ A {\bf 625},
 565 (1997).
\bibitem{RC.97} R.\ Rossignoli and N.\ Canosa, Phys.\ Lett.\ B {\bf 394},
 242 (1997); N.\ Canosa and R.\ Rossignoli, Phys.\ Rev.\ C \ {\bf 56}, 791
 (1997); R.\ Rossignoli, N.\ Canosa, P.\ Ring, Phys.\ Rev.\ Lett.\ {\bf 80},
 1853 (1998).
 \bibitem{K.87}J.\ Kurmann, H.\ Thomas, and G.\ M\"uller, Physica A
 {\bf 112}, 235 (1982).
\bibitem{RCM.08} R.\ Rossignoli, N.\ Canosa, and J.M.\ Matera (submitted)
\end{thebibliography}
\end{document}